# On Optimizing Shared-ride Mobility Services with Walking Legs

**Zifan Wang** [a]
Email: zifanw9@uci.edu
ORCiD: 0000-0002-3458-3748

**Michael Hyland** [b, c]
*Corresponding Author*
Phone: (949) 824-5084
Email: hylandm@uci.edu
ORCiD: 0000-0001-8394-8064

**Younghun Bahk** [b, c]
Email: ybahk@uci.edu
ORCiD: 0000-0001-5233-1563

**Navjyoth J.S. Sarma** [b, c]
Email: nsarma.js@uci.edu
ORCiD: 0000-0002-1304-0162

[a] *RMSI North America Inc.*
[b] *Department of Civil and Environmental Engineering, University of California, Irvine*
[c] *Institute of Transportation Studies, University of California, Irvine*

January 26, 2022

# Abstract

Shared-ride mobility services that incorporate traveler walking legs aim to reduce vehicle-kilometers-travelled (VKT), vehicle-hours-travelled (VHT), request rejections, fleet size, or some combination of these factors, compared to door-to-door (D2D) shared-ride services. This paper provides a review of shared-ride services with walking legs (SRSWL), particularly the studies in the literature that model the operational problem(s) associated with SRSWL. The paper describes the operational and societal benefits of SRSWL as well as compares the SRSWL to circuitous D2D shared-ride services, ride-hailing services, and fixed route transit services, in terms of VKT and traveler walking distance. The paper then delineates the operational subproblems associated with the SRSWL and discusses their computational complexity. Additionally, the review classifies configurations of SRSWL based on flexibility in assigning travelers to pickup and drop-off locations. The paper also discusses four modelling challenges: short-distance person trips, drop-off location choice for a vehicle's last remaining passenger, allowing vehicles to wait for travelers at pickup locations, and simultaneously reducing VHT/VKT and improving customer service quality relative to D2D shared-ride services. The review paper concludes by discussing the most critical areas of future research related to SRSWL.

# 1 Introduction

## 1.1 Background on Shared Mobility Services

Shared-ride services pool together multiple users, who are traveling in similar directions around the same time, into one vehicle. Prominent real-world examples of shared-ride services include UberPool and Lyft Shared, which are mobility-on-demand (MOD) shared-ride services offered by transportation network companies (TNCs) Uber and Lyft, respectively. Various transit agencies across the country operate (or contract others to provide) shared-ride services, often called micro-transit.

The shared-ride services described in the previous paragraph involve dedicated vehicles and drivers who provide service to customers. According to Shaheen & Cohen (2019), who provide a detailed taxonomy of shared-ride services, this type of shared-ride service is also known as ride-splitting. Another type of shared-ride service is conventional ridesharing/carpooling, in which drivers with their own trip origins and destinations pick up and drop off other travelers en-route from their own origin to destination. The current paper focuses on ride-splitting, but most insights are relevant to conventional ridesharing. As such, this paper uses the term shared-ride services for the general case and ride-splitting and ridesharing for specific cases.

By simultaneously serving multiple requests with one vehicle, shared-ride services can reduce vehicle-kilometers-travelled (VKT) and vehicle-hours-travelled (VHT) compared with a non-shared-ride taxi-like MOD service (i.e., ride-hailing). The reduction in VKT and VHT from shared-ride services may bring significant societal benefits such as reductions in vehicle-related air pollution (McCubbin and Delucchi, 1999; Zhang and Batterman, 2013) and energy consumption (Zhang et al., 2019; Zhou et al., 2016), as well as traffic congestion relief (Alisoltani et al., 2021).

Real-world ride-splitting services such as UberPool, Lyft Shared, and most conventional ridesharing applications involve a door-to-door (D2D) scheme. Under a D2D scheme, the vehicle always picks up and drops off requests at their requested origins and destinations, respectively. This often involves the vehicle taking circuitous routes, particularly around pickup (PU) and drop-off (DO) locations (or PUDO locations). Such inefficiencies may lead to unnecessary VHT/VKT for the vehicle fleet, limiting the benefits of shared-ride services.

One strategy to further reduce VHT and VKT is to require, request, or incentivize travelers to walk a short distance to/from nearby PU/DO locations that are convenient for the vehicle operator. This paper refers to such a service as a shared-ride service with walking legs (SRSWL)—a service that involves travelers being picked up and dropped off at alternative locations that are within a few blocks from their origin and destination locations. Ideally, the short walking trips for travelers can increase the directness of SRSWL vehicle routes compared with a D2D service.

Conventional bus transit routes, which are typically quite direct, can be considered as the simplest form of a shared-ride service that involves traveler walking legs. However, public transit typically involves fixed schedules and routes, which are less flexible than MOD ride-hailing and ride-splitting services, and the SRSWL discussed in this paper.

TNCs recently started experimenting with SRSWL. Around early 2018, Uber launched a service known as Express Pool that involves short walk legs (Lo and Morseman, 2018). In 2019, Lyft rolled out a similar service known as Shared Saver.

Figure 1 illustrates idealized vehicle routes for the four modalities discussed previously: fixed-route bus transit, SRSWL, D2D shared-ride service, and ride-hailing. The fixed-route transit service and the SRSWL require lower VKT and VHT, thereby decreasing congestion, energy consumption, and emissions as well as operational costs compared to D2D shared-ride services and ride-hailing. The lower operational costs may also permit lower prices for riders. On the other hand, D2D shared-ride services and ride-hailing only require travelers to walk very short distances, if at all—a valuable convenience. Therefore, there is a theoretical trade-off between societal goals and operational efficiency on one side and customer service quality on the other, in terms of the four modalities. Confirming and quantifying such a theoretical trade-off, understanding the service quality implications (e.g., waiting time, total trip time) of each modality, and designing good service configurations for each modality are important research problems with immediate relevance to TNCs, transit agencies, transportation planners, policymakers, regulatory agencies, and travelers.

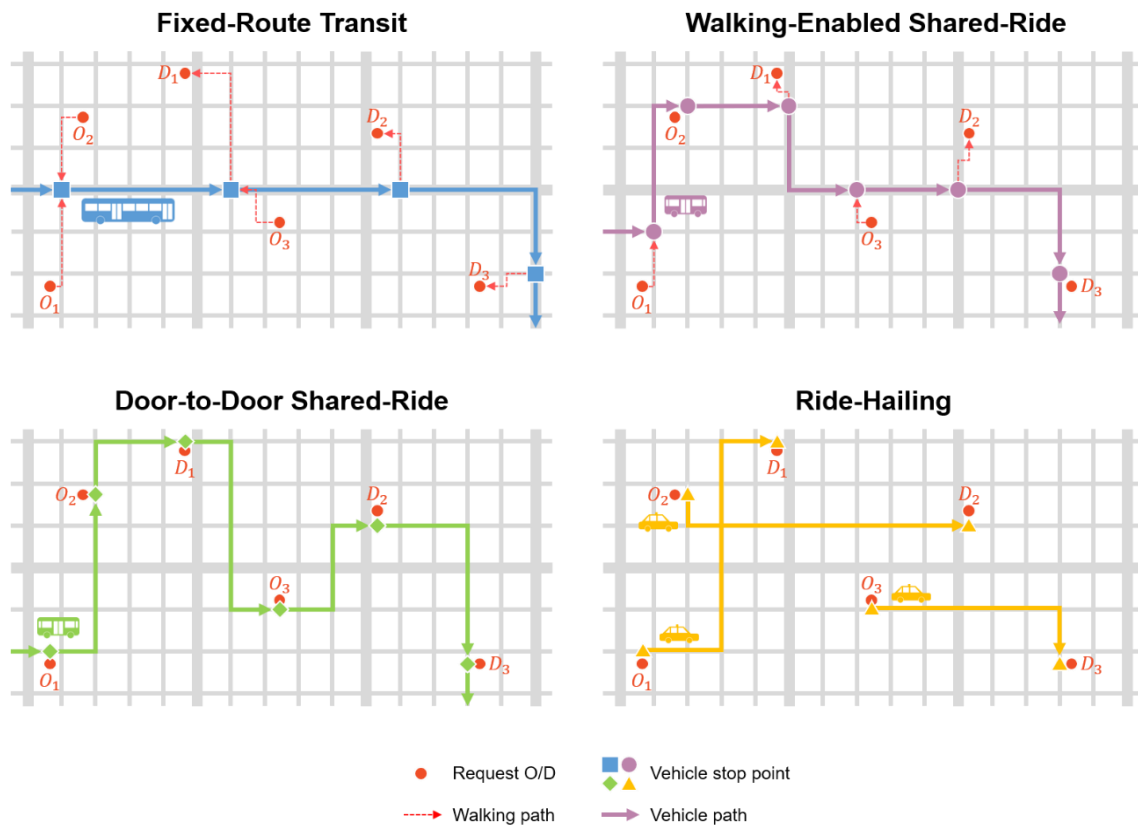

**Figure 1 Illustrative Routes for Four Shared Mobility Modalities**

## 1.2 Motivation for Review Paper

Models and quantitative analysis techniques offer an inexpensive means of assessing the benefits and disbenefits of the modalities in Figure 1, compared to operating variants of each service in the field. Many studies in the literature analyze fixed-route transit, ride-hailing, and D2D shared-ride services. Moreover, while the research topic has only arisen relatively recently, since 2014 around fifteen studies have investigated SRSWL from an operational perspective. While this is notably fewer than the number of studies investigating the other three modalities in Figure 1, there is still value in providing a critical review of the SRSWL literature and identifying future research directions, especially given the potential benefits of SRSWL.

Based on the authors' review of SRSWL studies in the literature, and the authors' own research in the area (Wang, 2021), researchers would benefit from a critical synthesis of the existing literature and an identification of the most critical future research questions. Despite significant advances in terms of tackling the SRSWL operational problem in recent years, it is a highly complex problem, and there are many outstanding modelling and algorithmic challenges remaining. Moreover, the relative benefits and disbenefits of SRSWL compared to the other three modalities in Figure 1 require further analysis. Similarly, there are several possible SRSWL configurations that require comparison and analysis.

## 1.3 Research Goals and Paper Structure

The overarching goal of this paper is to assess the existing SRSWL literature in order to provide insights to researchers, TNCs, transit agencies, planners, and policymakers and ultimately identify the most critical future research directions. To meet this goal, this paper reviews existing studies, delineates the components of the underlying operational problem (Section 2), discusses computational challenges (Section 2.2), classifies SRSWL by service configuration (Section 3), and identifies a series of potential modelling and algorithmic challenges worthy of further research (Section 4). The review focuses on models and algorithms related to the design, planning, and operation of SRSWL, along with studies analyzing the societal/system impacts of SRSWL. Section 5 concludes the paper by identifying and describing the most important future research directions.

## 2 Shared-ride Service with Walking Legs Problem (SRSWLP)

## 2.1 Problem Statement and Problem Context

The shared-ride service with walking legs problem (SRSWLP) can be generically defined as follows. Consider a SRSWL service provider with a vehicle fleet $V$, indexed by $v \in V$ with $|V|$ vehicles, that is responsible for providing mobility service to a group of travelers $R$, indexed by $r \in R$ and of size $|R|$, who request service over an analyzes period $T$, indexed by $t \in T$ with $|T|$ time intervals. Providing mobility service entails transporting each traveler $r \in R$ from their origin $o_r$ to their destination $d_r$. To do this, the service provider must (i) determine PU and DO locations for each traveler $r$, denoted $p_r$ and $q_r$, respectively; (ii) match/assign each traveler $r$ to a vehicle $v$; and (iii) determine an ordered sequence of request PUs and DOs for each vehicle $v$, denoted $P_v = \{p_{r1}, p_{r2}, q_{r2}, p_{r3}, q_{r1}, q_{r3} ...\}$ These are the three main subproblems associated with this general definition of the SRSWLP, denoted, PUDO location selection, request-vehicle assignment, and PUDO stop sequencing, respectively. However, the core of the SRSWLP—and what makes it a distinct class of vehicle routing—is the presence of the PUDO location selection subproblem. Hence, this paper focuses on the interaction between the PUDO location selection subproblem and the other two subproblems.

(The D2D shared-ride service and ride-hailing service operational problems do not incorporate the PUDO location selection subproblem because the PUDO locations are the traveler's exact origin ($p_r = o_r$) and destination ($q_r = d_r$), repectively. Moreover, the PUDO stop sequencing subproblem is highly simplified for ride-hailing because a traveler's drop-off must immediately succeed the same traveler's pickup in $P_v$.)

In addition to the three main SRSWLP subproblems, Hyland, Yang, & Sarma (2021) also identify network pathfinding as an additional subproblem for D2D shared-ride service case that is applicable to the SRSWLP. However, previous research on the SRSWLP does not explicitly consider this subproblem, rather existing studies assume vehicles always travel on the shortest path between PUDO locations. Additionally, for stochastic dynamic problem instances, some studies in the D2D ride-splitting literature (e.g. Alonso-mora et al., 2018) and most studies in the ride-hailing literature incorporate empty vehicle repositioning as a subproblem (Hörl et al., 2019; Hyland et al., 2019).

The service provider's objective in the SRSWLP, or any MOD service, may be to minimize operational costs, maximize profit, maximize service quality, minimize VKT, or some combination of these objectives (Hyland and Mahmassani, 2017). Additionally, the service provider faces constraints on the decision levers when solving the SRSWLP or its subproblems. These constraints include vehicle capacity constraints; time window constraints for request pickups, drop-offs, or destination arrivals; and service quality constraints related to walking distance/time, waiting time, and in-vehicle detour time. The time window and service quality constraints require an additional SRSWLP operational decision that is often combined with the PUDO stop sequencing subproblem—the vehicle scheduling problem. The vehicle scheduling subproblem involves determining the pickup time and drop-off time for every PUDO task in a vehicle $v$'s ordered sequence of requests, $P_v$.

This paper considers both deterministic (often called static) and stochastic dynamic versions of the SRSWLP. In the deterministic case, all information about the system is known before the analysis period begins. In the stochastic dynamic case, information about requests in $R$ are unknown prior to the analysis period $T$, rather the information reveals itself during the analysis period $T$. In the deterministic case, the SRSWLP can be solved once before the analysis period. In the stochastic dynamic case, sequential decisions based directly or indirectly on the three SRSWLP subproblems need to be made throughout the analysis period as new information reveals itself.

While it is possible to formulate the deterministic SRSWLP and its component subproblems in a single mathematical program, and to solve very small instances of this math program exactly, the various mathematical formulations of the SRSWLP in the literature all have limitations due to the computational challenges associated with the full/general SRSWLP. Balardino and Santos (2016) formulation does not include temporal components and all travelers have the same destination. Wang's (2021) formulation is only for a single vehicle setting. Zhao et al. (2018) assume a grid network and coarse temporal resolution. Zheng et al. (2019) formulation is tailored to a flexible-route transit setting. Finally, there are several studies that incorporate detailed mathematical expressions of components of the SRSWLP, but these expressions do not fully describe an integer-linear program (ILP).

The lack of a single coherent and comprehensive ILP formulation of the multi-vehicle SRSWLP reflects the complexity of the many variables, relationships, and possible modelling assumptions associated with the SRSWLP. Nevertheless, the authors of this paper believe a generalized ILP formulation of the multi-vehicle SRSWLP that incorporates all three subproblems would be of considerable value to the research community. Without this general formulation, it becomes difficult to compare solution algorithms and even specific service designs across studies.

Nevertheless, even with a generalized ILP formulation of the SRSWLP, the authors expect only very small problem instances to be computationally feasible with exact solution methods. Hence, as future sections describe, many researchers simplify the SRSWLP by limiting flexibility in terms of the PUDO location selection or tightening vehicle capacity constraints. Alternatively, researchers break the SRSWLP into subproblems and then solve the subproblems sequentially with the solution to one subproblem being the input to the next. The following section provides an overview of the subproblems and their computational complexity.

## 2.2 SRSWLP Subproblems and Their Computational Complexity

Figure 2 displays the three SRSWLP subproblems in a Venn diagram. At the center of the Venn diagram (Area 7) is the multi-vehicle SRSWLP that combines all three subproblems. Areas 4, 5, and 6 denote areas where two and only two of the subproblems overlap. Finally, Areas 1, 2, and 3, denote areas where one and only one of the subproblems is active. For Areas 1 through 6, the Venn diagram displays relevant routing problems in the literature.

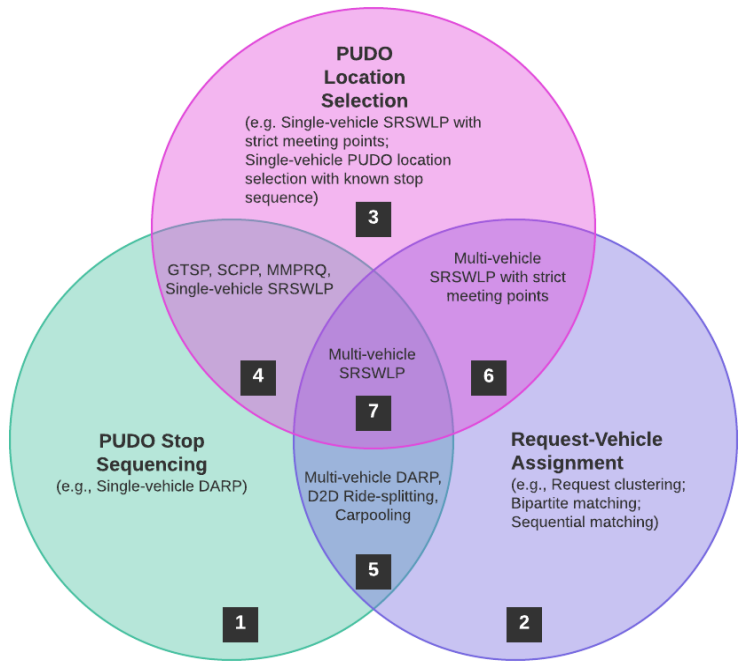

**Figure 2 Venn Diagram of SRSWL Subproblems**

The PUDO stop sequencing subproblem in Area 1 is a common problem in the academic literature and in practice; it is known in the passenger routing literature as the single-vehicle dial-a-ride problem (DARP) (Psaraftis, 1980). The problem is to optimize the sequence of PUDO locations of a single vehicle, given a set of requests. The problem often includes time-window, service quality, and capacity constraints. The DARP can be further decomposed into the pickup and delivery problem with time windows (PDPTW) (Desaulniers et al., 2002; Dumas et al., 1991) and the scheduling problem (Savelsbergh, 1992; Sexton and Bodin, 1985).

The single-vehicle DARP is NP-hard, although problem instances with a few dozen requests can be solved exactly by commercial solvers. Notably, any problem that includes the single-vehicle DARP (Areas 4, 5, and 7) as a subproblem will also be at least NP-hard.

The request-vehicle assignment subproblem in Area 2 is better known as the bi-partite matching problem or assignment problem. This problem involves assigning a set of agents to a set of tasks. Each agent-task combination has an associated cost or reward, and the objective is to assign agents to tasks such that the cumulative reward/cost is maximized/minimized. Despite being an integer program, the constraint matrix of the assignment problem is total unimodularity, meaning, relaxing the integrality constraints still produces optimal solutions with integer values. Hence, researchers can solve assignment problems with thousands of agents/vehicles and thousands of tasks/travelers in a few seconds.

The total unimodularity property of the assignment problem has been exploited by researchers solving dynamic vehicle routing problems in recent years. In the dynamic ride-hailing case, a common operational policy is to repeatedly solve instances of the request-vehicle assignment problem for available vehicles and unassigned requests every time interval (Dandl et al., 2019; Hörl et al., 2019; Hyland and Mahmassani, 2018; Maciejewski et al., 2016). In the dynamic D2D shared-ride case, a similar operational policy is used in which only one request can be inserted into a vehicle at each decision epoch (Hyland and Mahmassani, 2020; Simonetto et al., 2019), despite the service itself not having this limitation.

The PUDO location selection subproblem in Area 3 arises in SRSWL but is not found in most of the routing literature, as requests are typically assumed to specify their own PU and DO locations. Fielbaum (2021) studies the PUDO location selection subproblem, and proposes an algorithm relevant to the SRSWL no meeting points service configuration described in Section 3.3. Additionally, the single-vehicle variant of a strict meeting points service configuration described in Section 3.1 is another example of the pure PUDO location selection subproblem.

Area 4 displays problems that incorporate both PUDO stop sequencing and PUDO location selection. Example problems in Area 4 are the generalized travelling salesman problem (GTSP) (Laporte et al., 1987; Rice and Tsotras, 2013), the shortest covering path problem (SCPP) (Current et al., 1984; Niblett and Church, 2014), or the optimal multi-meeting-point route query (MMPRQ) problem (Li et al., 2016). Certain *single-vehicle* SRSWLP also combine these two subproblems. To be precise, such single-vehicle SRSWLP may include certain variants of the flex-route transit problems that allow requests to be picked up or dropped off at locations not necessarily their own origins/destinations or any of the mandatory checkpoints/stops (Malucelli et al., 1999; Qiu et al., 2014; Zheng et al., 2019), as well as the single-vehicle SRSWL formulation presented in Wang (2021). The GTSP, SCPP and MMPRQ are all NP-hard problem; hence, so is the single-vehicle SRSWLP.

The GTSP and the SCPP both aim to find the minimum-cost path that visits exactly one vertex in each cluster of vertices, where each vertex is an abstraction of a real-world location (e.g., a candidate PUDO location associated with a request), and a cluster captures one or more vertices corresponding to real-world locations that are within the specified walking range from/to a request's origin/destination. The GTSP typically assumes a real-world location may correspond to multiple vertices on the graph, with each vertex belonging to one cluster (Laporte et al., 1987; Rice and Tsotras, 2013). Conversely, the SCPP typically assumes only one vertex for each real-world location, and the vertex may belong to different clusters (Current et al., 1984; Niblett and Church, 2014). Unfortunately, neither the GTSP nor the SCPP standard formulations capture precedence relations (i.e., picking up before dropping off), time windows, or vehicle capacities, which are important considerations in a SRSWLP.

Area 5 combines PUDO stop sequencing and request-vehicle assignment. This is a common routing problem in practice and in the passenger transportation literature the problem is known as the multi-vehicle DARP (Cordeau, 2006; Ropke et al., 2007) or the multi-vehicle PDPTW. The multi-vehicle DARP is a known NP-hard problem. Area 5 incorporates the D2D shared-ride service problem.

Area 6 combines PUDO location selection and request-vehicle assignment but does not incorporate PUDO stop sequencing. This is a very uncommon problem in the routing literature, as stop sequencing is often an integral part of both request-vehicle assignment and PUDO location selection. However, this review paper does identify one such case—the strict meeting points SRSWL configuration reviewed in Section 3.1. In the strict meeting points SRSWL configuration, the decision problem involves determining a single PU location and a single DO location for each shared-ride vehicle and assigning requests to a single vehicle. The computational complexity of this problem grows with the number of feasible PU and DO locations for the vehicles, and with the number of vehicles and requests.

While solving a D2D shared-ride service problem is already computationally challenging because the underlying multi-vehicle DARP (Area 5) is NP-hard, adding walking legs introduces another level of complexity. The additional complexity is a function of the flexibility provided in the SRSWL for PUDO locations. As the number of feasible PUDO locations (or nodes) in the network (graph) for each request increase, so does the complexity of the SRSWLP.

The SRSWLP computational challenges have led researchers to make simplifying assumptions about the SRSWL itself, such as single-vehicle problems, restricting the number of intermediate stops in a vehicle sequence, all riders have the same destination, and restrictive locations for candidate PUDOs. Section 3 explores three different service configurations and their underlying assumptions and illustrates that certain service configurations that are more flexible have larger solution spaces and are computationally more challenging. But first, the next subsection reviews different methods to solve the SRSWLP and its subproblems sequentially.

## 2.3  Sequentially Solving the SRSWLP

This section covers solution approaches to solve the DARP (i.e., D2D shared-ride service problem) sequentially and then to solve the SRSWLP sequentially. Sequential approaches for the DARP include first clustering requests and then assigning and routing individual vehicles to serve each cluster of requests. This cluster first, route second method is quite common for deterministic problems (Bodin and Sexton, 1986; Borndörfer et al., 1999). Alternatively, studies first solve the single-vehicle DARP for many combinations of multiple requests and vehicles in order to determine the cost/reward associated with each potential match, and then second solve a matching problem based on the combinations of requests and vehicles and their associated costs/rewards (Santi et al., 2014). Alonso-Mora, Samaranayake, Wallar, Frazzoli, & Rus (2017) apply this method in a dynamic setting for a D2D shared-ride service via batching open requests and available vehicles every time interval (e.g. 30 sec). Interested readers should refer to Cordeau & Laporte (2007) and Mourad et al. (2019) for a comprehensive review of the DARP and D2D shared-ride mobility problems and solution algorithms.

In the SRSWLP literature, the PUDO location selection subproblem is often solved after (or along with) PUDO stop sequencing and before request-vehicle assignment (Fielbaum et al., 2021; Li et al., 2020; Lyu et al., 2019). Thus, for many request-vehicle combinations, it is necessary to solve a single-vehicle PUDO stop sequencing and location problem. Other studies (Araldo et al., 2019; Gökay et al., 2019; Gurumurthy and Kockelman, 2020) implicitly solve the PUDO location selection subproblem immediately upon receiving a traveler request, i.e., once a traveler makes a request the service provider selects the traveler's PU and DO locations (usually via a rule-based approach) before determining the new PUDO stop sequence and request-vehicle assignment. A third strategy, not yet developed in the literature, is to first solve the multi-vehicle DARP assuming the PU and DO locations are the origins and destinations of requests, respectively, and then second adjust PU and DO locations after request-vehicle assignment decisions are made.

The best sequencing of SRSWLP subproblems for stochastic dynamic and deterministic problems is an open research question that no existing studies address directly. While the answer is unlikely straightforward, even after conditioning on important problem parameters because the quality of solutions will depend on the implementation details of each sequencing method, comparisons are still likely to provide valuable insights into the SRSWLP itself.

## 3  Service Configuration Categorization

This section categorizes SRSWL in terms of one important service design dimension referred to as *service configuration*. The service configuration dimension captures the relative flexibility in assigning individual travelers to PUDO locations.

Broadly, the three types of service configurations found in the literature include: strict meeting points, relaxed meeting points, and no meeting points. The strict meeting points approach requires all travelers sharing the same vehicle to walk from their respective origins to one common PU location and to walk from one common DO location to their respective destinations. The relaxed meeting points approach involves algorithmic mechanisms that promote, but do not require, the sharing of common PU and/or DO locations among multiple travelers. The no meeting points approach, as the name suggests, makes little-to-no effort to assign multiple requests to common PU and or DO locations. Table 1 summarizes the key elements distinguishing the three service configurations, and it also includes examples of studies in the literature that address each service configurations.

The next three subsections detail each of these service configurations and review the studies related to each configuration. The fourth subsection outlines gaps in assessing the relative advantages and disadvantages of each configuration under a variety of scenarios.

**Table 1. Summary of Service Configurations for SRSWL**

|  | *Service Configurations* | | |
| --- | --- | --- | --- |
|  | **Strict Meeting Points** | **Relaxed Meeting Points** | **No Meeting Points** |
| **Is sharing of common PU or DO location among multiple travelers promoted?** | Yes | Yes | No |
| **Are multiple PU and DO locations possible?** | No | Yes | Yes |
| **Are drivers typically assumed to have their own destinations?** | Yes | No | No |
| **Examples** | Single-vehicle deterministic: Aissat and Oulamara (2014), Czioska et al. (2017b)<br><br>Multi-vehicle deterministic: Stiglic et al. (2015), Czioska et al. (2017a)<br><br>Equilibrium analysis: Yan et al. (2020) | Single-vehicle deterministic: Malucelli et al. (1999), Qiu et al. (2014)<br><br>Multi-vehicle deterministic: Martínez et al. (2014), Czioska et al. (2019), Miklas-Kalczynska and Kalczynski (2020)<br><br>Multi-vehicle stochastic dynamic: Gökay et al. (2019), Araldo et al. (2019), Gurumurthy and Kockelman (2020) | Single-vehicle deterministic: Zheng et al. (2019)<br><br>Multi-vehicle deterministic: Balardino and Santos (2016)<br><br>Multi-vehicle stochastic dynamic: Lyu et al. (2019), Li et al. (2020), Fielbaum et al. (2021) |

## 3.1 Strict Meeting Points Service Configuration

Studies utilizing a strict meeting point service configuration (Aissat and Oulamara, 2014; Czioska et al., 2017b, 2017a; Stiglic et al., 2015) typically assume drivers have their own origins and destinations and are willing to take riders along the way (i.e., conventional ridesharing/carpooling). Nevertheless, it is possible to adopt strict meeting points service configurations in the ride-splitting case; for example, Lyu et al. (2019) compare the performance of a strict meeting points service configuration with a no meeting points service configuration.

Yan et al. (2020) demonstrate the merits of SRSWL with a strict meeting points configuration by performing an equilibrium analysis. Their study demonstrates the superiority of a strict meeting points service configuration with variable rider waiting time in terms of improving the rider and driver experience in case of demand spikes, compared with the conventional strategy of dynamic pricing.

Strict meeting points configurations are ideal from a driver's perspective because they entail the fewest stops per driver, of the three configurations. However, a strict assumption about meeting points confines the process of picking up and dropping off riders to relatively small regions (i.e., the meeting point must be within the walking range of all riders sharing the vehicle) and limits the potential for matches between drivers and riders.

The feasibility of a SRSWL with strict meeting points depends on the region's demand rate. If the demand rate is high enough, drivers should be able to find riders quite often, and riders should be able to find rides, without having to walk long distances. However, if the demand rate is relatively low, riders either need to walk long distances to find meeting points where they overlap with other riders and drivers' routes, or they do not get matched frequently.

### 3.2 Relaxed Meeting Points Service Configuration

SRSWL with relaxed meeting points emphasize reducing intermediate stops (Araldo et al., 2019; Czioska et al., 2019; Gökay et al., 2019; Gurumurthy and Kockelman, 2020; Martínez et al., 2014; Miklas-Kalczynska and Kalczynski, 2020), but this configuration does not require a single PU location and/or a single DO location for riders. With relaxed meeting points, one or more intermediate PUDO locations are allowed, and multiple PUDO locations for serving multiple requests are possible. However, these studies often include algorithmic mechanisms that promote the sharing of a common PU or DO location among multiple travelers. Studies with relaxed meeting points service configurations rarely assume that drivers have their own destinations (a special case is Miklas-Kalczynska & Kalczynski (2020)), rather, most relaxed meeting point approach studies focus on ride-splitting services.

Some papers with relaxed meeting points employ solution techniques that involve clustering known requests first and then determining the meeting point for each cluster second (Czioska et al., 2019; Martínez et al., 2014). Alternatively, relaxed meeting points service configurations may utilize techniques that restrict the candidate PUDO locations to a small subset of road nodes where requests may originate or terminate, and travelers are automatically assigned to the PUDO locations closest to their origin and destination, respectively. Once the PUDO locations for a request are determined, the operational problem becomes a conventional DARP. Strategies for restricting the candidate PUDO locations in the study area may include limiting candidate PUDOs to (i) the PUDO locations of currently in-service requests plus the cluster centroids of historical requests (Gökay et al., 2019), (ii) road junctions with regular stop spacing in a grid network (Araldo et al., 2019), or (iii) the cluster centroids of locations where requests might originate or terminate (Gurumurthy and Kockelman, 2020).

Additionally, flex-route transit services that permit travelers to walk to/from nearby stops corresponding to other accepted traveler requests' origin or destination (Malucelli et al., 1999; Qiu et al., 2014) can also be viewed as having a relaxed meeting points service configuration. Nevertheless, flex-route transit services that only provide a D2D service for travelers with requested PUDO locations not at checkpoints do not fit the three distinct service configurations outlined in this section, since such services only involve the PUDO stop sequencing subproblem (it does not deal with the PUDO location selection subproblem).

Other relaxed meeting points service configurations exist. For example, researchers may choose to adopt a hybrid shared-ride scheme in which all travelers sharing the same vehicle could be either served by the D2D scheme or served by the strict meeting points scheme (Miklas-Kalczynska and Kalczynski, 2020).

## 3.3 No Meeting Points Service Configuration

The no meeting points service configuration incorporates walking in shared-ride services without focusing on the reduction in the number of intermediate stops as a primary objective or constraint. Assigning multiple travelers to a common PUDO location is possible but would be driven by other objectives (e.g., minimizing VKT or VHT), or constraints/problem setting (e.g., only one feasible PUDO node nearby multiple requests).

No meeting points configurations tend to be flexible in assigning travelers to PUDO locations. All the candidate PUDO locations (e.g., road junctions) within the walking range from/to a request origin/destination might be considered.

The high flexibility in PUDO location selection entails a large solution space that is likely to contain better solutions in terms of service quality and operational efficiency, compared with the other two configurations. However, the large solution space increases the computational complexity of the problem, preventing exact algorithms from finding optimal solutions. Hence, the literature includes various heuristic techniques to obtain a reasonable solution within a reasonable time.

Table 2 summarizes the key features of studies that utilize no meeting points service configurations. Some studies (Balardino and Santos, 2016; Zheng et al., 2019) propose variants of the SCPP or GTSP formulations for the carpooling or flex-route transit applications. Furthermore, some recent works (Fielbaum et al., 2021; Li et al., 2020; Lyu et al., 2019) compare SRSWL to D2D ride-splitting services.

Notably, Lyu et al. (2019) maintain a waiting queue for unassigned requests, attempt to match a new request with requests in the waiting queue, determine the least cost set of requests to be pooled in one vehicle, and assign the closest idle vehicle to the first stop in the stop sequence corresponding to the set of least cost requests. N. Li et al. (2020) evaluate each candidate vehicle for each new request sequentially; test inserting the new PU and DO stops associated with the new request after each of the existing stops of non-idle candidate vehicles; select the PUDO location based on the least-cost D2D stop sequence; and choose the vehicle associated with the least-cost stop sequence and the adjusted PUDO locations. Fielbaum et al. (2021) batch new requests that are received in a time interval and solve an assignment problem that may assign multiple requests to the same vehicle for a time interval.

**Table 2. Key features of the five studies that utilize no meeting points service configurations**

|  | **Balardino & Santos (2016)** | **Zheng et al. (2019)** | **Lyu et al. (2019)** | **Li et al. (2020)** | **Fielbaum et al. (2021)** |
|---|---|---|---|---|---|
| **Demand Information** | Demand known in advance | Demand known in advance | Demand revealed over time | Demand revealed over time | Demand revealed over time |
| **Number of Vehicles** | Multiple | Single | Multiple | Multiple | Multiple |
| **Request-Vehicle Assignment** | Handled as a part of the ILP and the iterated local search heuristic | N/A | Sequential Assignment | Sequential Assignment | Batch matching (multiple requests to one vehicle possible) |
| **Problem context** | Carpooling | Flex-route transit | Ride-splitting | Ride-splitting | Ride-splitting |
| **Vehicle Capacity** | 4 | N/A (no capacity constraints) | 4 | 20 | 4 or 10 |
| **Will vehicle wait at PU location for travelers to arrive?** | N/A (not time-dependent) | N/A (travelers arrive first) | Yes | No | No |
| **Can requests be added to non-idle vehicles?** | N/A | N/A | No | Yes | Yes |

## 3.4   Choice of Service Configuration

This subsection identifies potential future research directions. As discussed in Section 3.2 and 3.3, both relaxed meeting points and no meeting points configurations have the theoretical potential to increase sharing, operational efficiency, and service quality relative to the strict meeting points approach. However, only a handful studies (Lyu et al., 2019; Miklas-Kalczynska and Kalczynski, 2020) analyze and provide evidence showing the superiority of the relaxed and no meeting points service configurations. Furthermore, it appears that no study provides a comparison between the relaxed meeting points approach and the no meeting points.

There are two major future research directions to address this gap. The first assumes a deterministic problem setting where information about all requests is known before the analysis period begins. A comparison can be done using the service configurations similar to the strict meeting points in Stiglic et al. (2015), the relaxed meeting points in Czioska et al. (2019) or Martínez et al. (2014), and the no meeting points in Balardino & Santos (2016).

The second direction assumes a stochastic dynamic setting. The relaxed meeting points approach in Araldo et al. (2019) or Gurumurthy & Kockelman (2020) should be compared with the no meeting points approach in Fielbaum et al. (2021) or Li et al. (2020). Additionally, fixed-route transit services can be introduced to the comparison study as well, similar to the comparison in Li et al. (2020).

The goal of these proposed comparison studies should not be to determine unequivocally the superior service configuration, but rather to assess the relative advantages and disadvantages of each configuration under a variety of scenarios.

## 4   Modelling and Algorithmic Challenges

This section discusses four modelling and algorithmic challenges that are unique to SRSWL. These modelling issues require careful consideration from TNCs, modelers, and researchers.

## 4.1 Treatment of Short-Distance Trips

When a traveler who requests SRSWL has an origin-to-destination distance less than the sum of the service's allowable pickup and drop-off distances, the service provider can deploy different strategies to accommodate this traveler. This section presents four cases (see Figure 3) that demonstrate the challenges involved in dealing with short-distance trips.

Given the vehicle's planned path and a new request's origin and destination as shown in Figure 3a, a naïve SRSWL service may prescribe a DO location that is the exact same as the PU location for the new request. Effectively, the vehicle may be instructed to pick up and drop off the traveler, but the traveler does not actually travel any distance inside the vehicle and eventually walks from the "PU/DO" location to her destination. Theoretically, this extreme but not unlikely case could lead to higher travel times for passengers already onboard who incur a delay as the vehicle detours to the short-distance traveler's PU/DO location. Additionally, the new traveler may have to wait for the vehicle to arrive at the PU location before continuing to walk to their destination, thereby increasing their waiting time. Therefore, algorithms should ensure such solutions to this edge case are made infeasible for short distance requests.

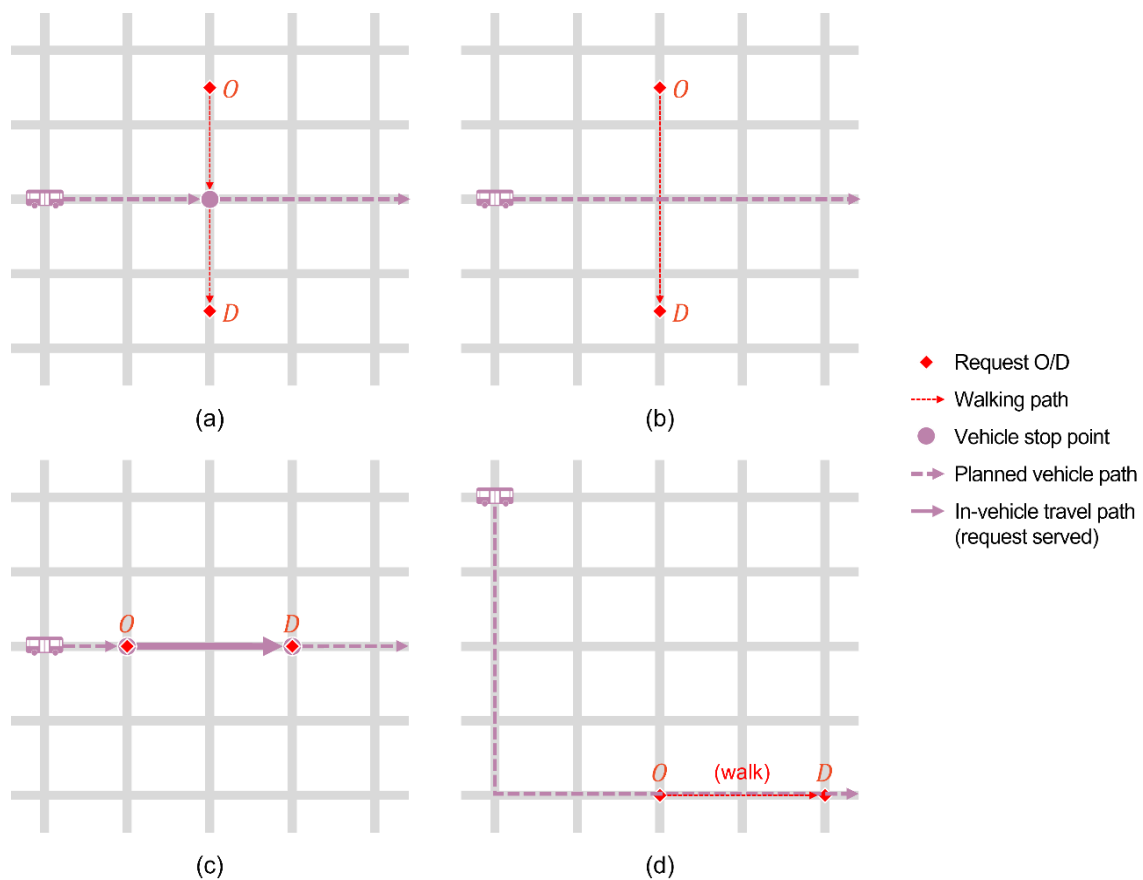

**Figure 3 Four possible extreme cases in short-distance trips: (a) the service does not provide any effective movement; (b) the travel is completed by walk without a service; (c) a vehicle provides efficient service without any walking required; and (d) the traveler would rather walk than wait for a distant vehicle.**

To avoid the problematic 'solution' shown in Figure 3a, one strategy is to always instruct the traveler to walk for the entire length of the trip. In this case, the traveler has zero wait time, and a vehicle does not need to detour to serve the traveler, as shown in Figure 3b.

Nevertheless, simply accommodating all requests for short-distance trips via walking may sometimes fail to realize the potential benefits associated with accommodating the trips via a shared-ride service. For example, Figure 3c displays a scenario in which the traveler's origin and destination are located on the vehicle's planned path, and the vehicle is close to the traveler's origin. In this case, providing a D2D service to the traveler would relieve the burden of walking and possibly reduce total travel time for the traveler, while the impact on other passengers' riding experience would be minimal.

However, if the scenario in Figure 3c changes such that the only candidate vehicle is far away from the new request's origin, as shown in Figure 3d, it might be faster—and perhaps better—for the new traveler to walk directly to their destination than wait for a vehicle to pick them up.

This last case demonstrates the need to model the interaction dynamics between the traveler and the service provider. In the real world, a traveler who requests the SRSWL for a short distance trip is likely comparing the trip with a walking-only trip. As such, advanced models can capture service provider's offer to the traveler and the traveler's acceptance or rejection of the offer, in which the offer includes both the service quality attributes (walk distance, wait time, in-vehicle travel time [IVTT], etc.) and price. In the case where the service provider offers the traveler service similar to Figure 3c (i.e., short wait time and D2D service) the request would likely accept; whereas a traveler receiving an offer consistent with Figure 3d may reject the offer and walk depending on the traveler's relative disutility associated with walking and waiting.

Yu & Hyland (2020) present a model that captures a request's response time and decision to reject or accept a service provider's initial and final offer in the context of a ride-hailing service. Additionally, Erdmann, Dandl, & Bogenberger (2021) model sending initial heuristic-based wait time offers/estimates to requests, waiting for requests to respond to the initial offer, and then assigning a vehicle to the traveler using a more advanced assignment process, for a ride-hailing service. These modelling components can and should be embedded into models for SRSWL.

## 4.2 Drop-off Location for a Vehicle's Last Assigned Drop-Off

In a dynamic problem setting, idle vehicles are assigned to serve new requests. A question facing the service provider in the no meeting points SRSWL is: should the idle vehicle always offer a D2D service for new requests or should the request be assigned to walk to a PU location from the origin and/or from a DO location to their destination? This section focuses on the DO location to destination question, as it pertains not just to idle vehicles but to all assigned vehicles who have a last planned drop-off.

Figure 4 displays the results of three separate cost/control functions/strategies in terms of the selection of the last DO location. The figure shows that under each control function, the vehicle route and new request DO location are different. A pure D2D service (case a) involves dropping off the request at its destination location. Moreover, minimizing VHT/VKT exclusively (case c) versus minimizing (or at least considering) request walking distance alongside VHT/VKT (case b) produces two very different vehicle routes and DO locations.

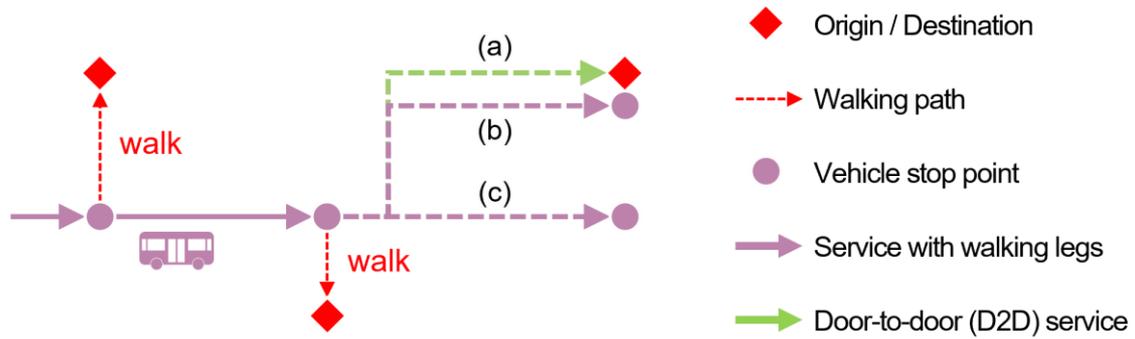

**Figure 4 Drop-off location for a vehicle's last assigned drop-off: (a) D2D for the last drop-off, (b) minimize walking distance/time and VHT, and (c) minimize VHT exclusively.**

Assuming the service provider's primary objective is to minimize VHT or VKT while the traveler's walking time or total trip time has little-to-no weight when assigning PUDO locations, the vehicle would drop off its last traveler as soon as it reaches a DO location within the walking range of the traveler's destination. Moreover, in this case, as the service's maximum drop-off walking distance increases, so too will the traveler's walking distance in the scenario with the VHT/VKT primary objective. Even worse, it's conceivable that a new request is added to an existing vehicle's PUDO sequence that would require the vehicle to bypass the originally assigned traveler's destination location. Such circumstances should be avoided.

One simple strategy to mitigate the biases of minimizing VHT or VHT in selecting the DO location for the last stop in a vehicle stop sequence is to always drop off the last remaining passenger at its destination location, like in Lyu et al. (2019). A justification for this strategy is that, by definition, (at the timestep under consideration) there will be no other passenger onboard the vehicle when the traveler in question is dropped off, so it is unnecessary to require the traveler to walk in an effort to reduce vehicle detours. However, this ignores the downstream impacts of decisions made at the current time step.

Alternatively, researchers might consider giving a higher weight to service quality factors like total travel time and walking time when choosing PUDO location for travelers in a multi-criteria approach. However, such a strategy has a potential bias of favoring a D2D drop-off service. For example, if the increases in both walking time and vehicle driving time are used to select PUDO locations, and they are weighed equally and measured in the same unit (e.g., seconds), the selected DO locations are likely to be the requests' destinations because it usually takes much less time to traverse a road segment via car than foot. Such a bias not only applies to the last stop in vehicle stop sequence but to any DO location in vehicle stop sequence.

The discussion in this section also indicates a need for researchers to develop anticipatory and non-myopic approaches rather than exclusively greedy/myopic approaches for solving the dynamic SRSWLP and its subproblems. None of the control functions or operational policies reviewed in this section take into consideration the downstream impacts of decisions made at the current time step. This contrasts with the literature on ride-hailing and D2D shared-ride services that incorporate demand forecasts into repositioning and assignment decisions (Dandl et al., 2019), employ approximate dynamic program techniques (Al-Kanj et al., 2020) and deep reinforcement learning techniques (Kullman et al., 2021) that explicitly consider the evolution of exogenous and endogenous factors and the downstream impacts of decisions made at the current time.

### 4.3 Vehicles Waiting for Travelers at Pickup Locations

For a shared-ride vehicle to pick up a request both the vehicle and the traveler must be at the PU location at the same time. While this is an obvious statement, it raises questions from a service design perspective. The one this section focuses on is: should vehicles wait at the PU location for travelers to arrive and, if so, for how long should they wait? In shared-ride services without walking legs, this issue is not as pertinent as the request should already be at its origin/PU location when it makes the request; however, Hyland & Mahmassani (2020) do analyze the case where some travelers make D2D shared-ride vehicles wait at their PU locations. Conventional transit (e.g., bus) service vehicles only wait a few seconds, if at all, at a bus stop/PU location for travelers to arrive.

Two recent studies on dynamic SRSWLPs (Fielbaum et al., 2021; Li et al., 2020) select a PU location that ensures the traveler arrives before the vehicle arrives. In other words, they explicitly forbid the vehicle waiting curbside for travelers. From a service operation perspective, this approach prevents the curbside parking problem associated with allowing vehicles to wait for travelers. Additionally, travelers who are onboard vehicles do not have to wait curbside for other travelers to walk to their PU location.

Alternatively, researchers may choose to allow vehicles to wait curbside for travelers to walk to their assigned PU locations, though this vehicle waiting time needs to be properly accounted for in the appropriate decision variables, constraints, and objective function terms (Lyu et al., 2019). If the PU location selection criterion focuses on reducing the VKT or VHT, the operator may favor PU locations that involve the vehicle waiting for the traveler, under the condition that this location reduces vehicle detour and therefore VHT and VKT.

However, such a shorter vehicle detour sometimes may occur at the expense of longer walking times and total trip times since walking is much slower than driving. If the PU location selection criterion focuses on reducing total trip time that includes walking and waiting, there is a possibility that the resultant total trip time and vehicle driving time would be shorter than the first alternative. Yet, none of the existing studies allow vehicles to wait for travelers while using a PU location selection criterion that gives a sufficiently large weight on walking and waiting; ideally future studies should provide more insights about the trade-off between forbidding and allowing vehicles to wait for travelers at PU locations.

### 4.4 Traveler Total Trip Time to Reach Destination

The existing literature consistently finds that enabling or requiring walking in shared-ride service can reduce VKT and VHT, even if the optimization framework does not explicitly focus on minimizing VHT or VKT (Czioska et al., 2019; Fielbaum et al., 2021; Lyu et al., 2019). However, compared with D2D shared-ride services, is it possible for a SRSWL to also reduce traveler total trip time when including waiting, walking, and IVTT?

Travelers' total trip time in SRSWL depends on two competing forces. The first force, which decreases total trip time, is the push towards more direct routes for vehicles resulting in significant decreases in traveler IVTT, relative to a D2D shared-ride service. The second force, which increases total trip time, is the need for travelers to walk from DO locations, and to a lesser extent, to PU locations. Walking is particularly time-consuming compared to vehicle-based travel, especially for the last-mile of a trip, where walking is not substituting for wait time like in the trip's first mile.

If walking time is not penalized sufficiently in constraints or objective function terms, the prescribed PUDO locations may entail a long walking time such that it cannot be fully offset by the reduction in IVTT. Fielbaum et al. (2021) give a large weight for walking in their cost function and show that the total trip time may decrease for SRSWL compared to D2D shared-ride service in their Manhattan case study. One scenario in Li et al. (2020) demonstrates that the total trip time may decrease if walking legs are only allowed for the origin to PU location, and not the DO location to destination. Conversely, Li et al. (2020) show that allowing/requiring both PU and DO walking legs generally does not decrease total trip time.

As mentioned in Section 4.2, assigning a large penalty for walking distance/time may bias solutions toward more D2D-like service, especially in the case of drop-offs. The results of Li et al. (2020) and Fielbaum et al. (2021) thus provide evidence supporting the importance of penalizing drop-off walking, if researchers want to reduce traveler total trip time in a SRSWL. It also should be noted that both Li et al. (2020) and Fielbaum et al. (2021) adopt methods that belong to a no meeting points approach, which provides ample flexibility in selecting optimal PUDO locations. The results of existing works that rely on a meeting points approach do not show potential to reduce traveler total trip time.

Interestingly, Fielbaum et al. (2021) also show that considering walking may still increase total trip time in three out of four scenarios in their toy network case studies, even though they consistently assign a relatively high penalty for walking and waiting. This finding probably reflects the fact that a short drop-off distance for one traveler may increase the detour time (i.e., extra IVTT) and waiting time of other travelers sharing the same ride, compared with encouraging a longer drop-off distance. The shorter the drop-off walking distance (and thus total trip time) of a given traveler, the longer the IVTT of other passengers onboard the vehicle as well as the waiting time of those travelers who are scheduled to be picked up after dropping off the given traveler. Such a trade-off complicates the effect of drop-off walking on the system-wide total trip time. The results of Fielbaum et al. (2021) suggest that a concentrated spatial distribution of demand and a network with a uniform block size may help reduce traveler total trip time. Nevertheless, future research should work to identify scenarios or combinations of input parameters where SRSWL simultaneously reduce traveler total trip time and VHT/VKT, compared with a D2D shared-ride service. Researchers should also work to develop control strategies that can reduce both VKT/VHT and traveler total trip time.

# 5 Conclusion

This paper presents an overview of the SRSWL operational problem, as well its component subproblems—PUDO location selection, request-vehicle assignment, and PUDO stop sequencing. The paper discusses the computational complexity associated with the three subproblems and their integration. This paper also identifies three major categories of service configurations for SRSWL—strict, relaxed and no meeting points configurations. Additionally, the paper identifies and discusses four modelling challenges that are unique to SRSWL. The remainder of this section itemizes and summarizes the most important future research directions identified in this study.

## 5.1 Formulating the Generalized Multi-vehicle SRSWLP

This paper notes that the literature is lacking a complete, non-simplified ILP formulation of the deterministic multi-vehicle SRSWLP. Most existing formulations include strong simplifying assumptions about SRSWL service designs that limit their generalizability. A complete ILP formulation of the multi-vehicle SRSWLP is likely to provide valuable insights into the complexity of the problem as well as allow for comparisons across service designs and solution algorithms.

The paper specifically identifies four modelling challenges associated with the SRSWLP, including, handling short-distance trips; DO location selection for a vehicle's last remaining passenger; allowing or disallowing vehicles to wait for travelers at PU locations; and the degree of emphasis in reducing traveler total trip time (rather than solely minimizing VHT). ILP formulations of the SRSWLP should include decision variables, constraints, and objective function terms to incorporate each of these four items.

## 5.2 Optimizing SRSWL

Optimizing SRSWL is the focus of the paper. Both the deterministic SRSWLP and the stochastic dynamic SRSWLP are computationally challenging problems, and the existing modelling and algorithmic developments related to SRSWLP are somewhat limited. Most studies develop a model and solution algorithm that they then use to perform systems analysis. While modelling and solving the SRSWLP is undoubtedly challenging due to the problem's inherent complexity, future research should develop models to significantly improve the operational performance of SRSWL and develop algorithmic strategies to address the problem's computational complexity.

In the case of the stochastic dynamic SRSWLP, none of the existing operational policies explicitly incorporate the downstream impacts of decisions made at the current decision interval (i.e., the operational policies are reactive and myopic rather than anticipatory), a significant shortcoming. The authors of this paper believe this is the most fruitful area of future research related to optimizing SRSWL. Operational policies based on reinforcement learning, approximate dynamic programming, cost-function approximations with optimized hyperparameters, and model predictive control have the potential to significantly improve SRSWL performance—see Powell (2019) for an overview of optimal policies for stochastic dynamic problems. The main challenge with implementing these advanced policies is the size of the decision space and state space in the Markov decision process framing, and the size of the decision variable vector in the control theory framing. Integrating PUDO location selection, request-vehicle assignment decisions, and PUDO stop sequencing (and scheduling) produces an extremely large decision space/vector.

Additionally, future research should compare different solution algorithms that sequence the three SRSWLP subproblems, for both deterministic and stochastic dynamic settings. The authors of this paper, in the stochastic dynamic case, are currently comparing (i) the sequence stops and assign vehicles to requests first, then adjust PUDO locations second approach to (ii) the sequence stops and choose PUDO locations first, then assign vehicles to requests second approach.

## 5.3 Comparing SRSWL to other Shared Mobility Modalities

Comparing SRSWL to other shared mobility modalities is another important area of future research. The introduction of this paper lays out the advantages and disadvantages of SRSWL relative to ride-hailing, D2D shared-ride services, and fixed-route transit. However, extensive analyzes of the relative benefits and disbenefits of SRSWL compared to the other three modalities is missing from the existing literature. It is critical to identify the circumstances wherein each modality performs effectively and ineffectively, and to understand the trade-offs between the modalities under these circumstances. Similarly, the existing literature is missing a comparison of the relaxed meeting points approach to the no meeting points approach.

## Acknowledgements

The second, third, and fourth authors of this paper received partial funding from the U.S. Department of Energy Vehicle Technologies Office under the Systems and Modeling for Accelerated Research in Transportation Mobility Laboratory Consortium, an initiative of the Energy Efficient Mobility Systems Program.

**Glossary of Acronyms**

| | |
|---|---|
| D2D | door-to-door |
| DARP | dial-a-ride problem |
| DO | drop-off (when referring to a drop-off location) |
| GTSP | generalized travelling salesman |
| ILP | integer-linear program |
| IVTT | in-vehicle travel time |
| MOD | mobility-on-demand |
| MMPRQ | multi-meeting-point route query |
| PDPTW | pickup and delivery problem with time windows |
| PU | pickup (when referring to a pickup location) |
| PUDO | pickup and/or drop-off |
| SCPP | shortest covering path problem |
| SRSWL | shared-ride service(s) with walking legs |
| SRSWLP | shared-ride service with walking legs problem |
| TNCs | transportation network companies |
| VHT | vehicle-hours-travelled |
| VKT | vehicle-kilometres-travelled |